\documentstyle[12pt]{article}
\def\ni{\noindent} 
\def\Dc{{\cal D}}
\def\no{\nonumber}
\begin{document}
\begin{center}
{\Large {\bf Canonical Transformations and Gauge Fixing in the Triplectic Quantization}}\\
\vspace{1cm} 

{\large Everton M. C. Abreu,  Nelson R. F. Braga \\ and Cresus F. L. Godinho}  \\
\vspace{1cm}

 Instituto de F\'\i sica, Universidade Federal  do Rio de Janeiro,\\
Caixa Postal 68528, 21945  Rio de Janeiro,
RJ, Brazil\\

\vspace{1cm}

\large April 8, 1998

\end{center}

\vspace{1cm}
\abstract
We show that the generators of  canonical transformations in the triplectic manifold   must satisfy constraints that have no parallel in the usual field antifield quantization.
A general form for these transformations is presented. Then we consider  gauge fixing by means of canonical transformations in this Sp(2) covariant scheme, finding a relation between generators and gauge fixing functions.
The existence of a wide class of solutions to this relation nicely reflects the large freedom of the gauge fixing process in the triplectic quantization.
Some solutions for the generators are discussed. 
Our results are then illustrated by the example of Yang Mills theory.  


\vspace{1cm}


\vspace{3cm}

\noindent everton@if.ufrj.br; braga@if.ufrj.br; godinho@if.ufrj.br

\vfill\eject
\section{Introduction}
The standard form of the field antifield or Batalin Vilkovisky (BV) Lagrangian quantization scheme \cite{BV1,BV2,HT} is based on  imposing explicit BRST invariance  of the Vacuum functional. 
In this approach, antighosts (as well as auxiliary fields)
do not enter the minimal set of fields. They show up just in trivial pairs that one must include in the theory in order to allow a standard gauge fixing procedure.

\bigskip

An alternative formulation for the BV quantization, where Sp(2) invariance is imposed, was presented in \cite{BLT1,BLT2,BLT3}. In this formulation one needs two kinds of antifields: $\phi_A^{\ast\,1}$  that correspond to sources of BRST transformations, $
\phi_A^{\ast\,2}$ corresponding to sources of the anti-BRST ones and also auxiliary fields ${\bar \phi}^{A}$ that generate the product of both kinds of transformations. In this formulation, gauge fixing corresponds to adding  a double (BRST times anti-BRS
T) variation of some bosonic functional to the original action and demands the introduction of two sets of auxiliary fields $\pi_A^{1}$ and $\pi_A^{2}$. 
In this scheme, extended invariance requires the existence of antighosts and auxiliary fields in the  minimal set of fields $\phi^A$.

\bigskip

A general formulation for gauge theory quantization with extended BRST invariance (BRST and anti-BRST) was presented in \cite{GH1,GH2,GH3}. In these articles, extended invariance is implemented by starting from a standard BRST cohomological approach, but 
duplicating the gauge generators and the ghost structure. The auxiliary fields show up naturally as ghosts for ghosts associated to the reducibility of the duplication process.  In particular, in reference \cite{GH3} the variables corresponding to the fie
lds $\pi_A^{1}$ and $\pi_A^{2}$ are introduced as the canonical conjugates to ${\bar \phi}^{A}$.  This approach has the important consequence of preserving  an important feature of the standard BV quantization: the anticanonical form of the coordinate spa
ce (of fields and antifields), where the variables show up in pairs that are conjugate with respect to the antibracket operation.
In the  extended version of \cite{BLT1,BLT2,BLT3} this anticanonical form was not present as it was based on two antibrackets (one for each of the antifields) that have essentially the same form as those of the standard BV quantization.
Therefore they were not sensible to the variables ${\bar \phi}^A $ , $\pi_A^{1}$ and $\pi_A^{2}$, that are necessary ingredients of a BRST extended formulation.

\bigskip

This idea of defining  an extended configuration space, in the case of the Sp(2) invariant quantization, with a completely anticanonical form  was recently put in general grounds in the so called triplectic quantization\cite{T1,T2,T3}.  In this formulatio
n one works with an extended version of the antibrackets, such that all of the field variables   will be accommodated in conjugated pairs.  

\bigskip

An important aspect of the field antifield quantization is that of canonical transformations\cite{BV1,BV3,H1,TPN,DJ,GPS}, that means, transformations that do not change the form of the antibrackets.
Gauge fixing in the  standard BV quantization  can be implemented by means of a canonical transformation to a gauge fixed basis\cite{DJ}. In this case there is a direct trivial relation between the generator of the canonical transformation and the corresp
onding gauge fixing fermion. In other words, the same gauge fixed action can be obtained either by gauge fixing in the standard way with some fermionic functional $\Psi$ or by performing a canonical  transformation generated by the same $\Psi$.

\bigskip

The question that we will address in this article
is: how do these features appear in the triplectic quantization? In other words, what is the general form of  transformations that are canonical with respect to both kinds of triplectic antibrackets and how can we use them to implement gauge  fixing?

\bigskip

The gauge fixing procedure in a formulation with  extended BRST invariance is more elaborated than in the standard BV quantization. There is a large freedom in choosing a gauge fixing action in the extended triplectic field antifield space. As  discussed 
in \cite{DJB}, one can not interpret gauge fixing as just the process of replacing antifields by some function of the fields.
We will see that this large freedom in the gauge fixing process will be reflected in the freedom in choosing canonical transformations that lead to some final gauge fixed action.
We will find a general relation between generators of canonical transformations and gauge fixing bosonic functions and  present  some solutions. 

\bigskip

The article is organized as follows: we will  briefly review in section (2)  some aspects of the extended BRST quantization and in section (3) the so called triplectic quantization. In section (4) we study the canonical transformations in this formalism.
In section (5) we discuss the relation between gauge fixing and canonical transformations. The Yang Mills theory  illustrates our results in section (6).
Finally section (7) contains some final remarks and conclusions.

\section{Extended BRST quantization}
We will here present a short review of the method of references \cite{GH1,GH2,GH3}.
One introduces an extended BRST operator $\delta \,=\,\delta_1\,+\,\delta_2$ where 
$\delta_1$ and $\delta_2$ are respectively the BRST and anti-BRST operators, that satisfies 

\begin{equation}
\label{NBRST}
\delta^2\,=\,0 
\end{equation}

\ni Associating to $\delta_1$ and $\delta_2$ ghost numbers 1 and -1 respectively,
equation (\ref{NBRST}) contains the full extended BRST algebra

\begin{equation}
\delta_1^2\,=\,0\,\,,\,\,\,\delta_2^2\,=\,0\,\,,\,\, \delta_1\,\delta_2\,+\,
\delta_2\,\delta_1\,=\,0
\end{equation}

One then follows the BRST approach of associating ghosts to gauge symmetry parameters but, in contrast to the standard procedure, one  duplicates the set of gauge generators and correspondingly also the set of ghosts, by introducing also the antighosts in
 a symmetric way. 
An interesting feature of this procedure is that the reducibility of such a trivial duplication of generators leads to the need of introduction of ghosts for ghosts, that will play the role of the standard auxiliary fields.

Then in order to accommodate this duplicated structure, instead of introducing  just the standard BRST  grading: the ghost number, let us introduce a bidegree $bigh\,=\,(gh_1\,,\,gh_2\,)\,$, that will allow the separation between the BRST and anti-BRST st
ructures. The bidegree is  related to the standard ghost number $gh$ and to the new ghost number $ngh$ defined on \cite{BLT1}, for some object $X$ (field, antifield,...) by

\begin{eqnarray}
gh (X) &=& gh_1 (X)\,-\,gh_2 (X)\nonumber\\
ngh (X) &=& gh_1 (X)\,+\,gh_2 (X)\,\,.
\end{eqnarray}

\noindent Classical fields will have bidegree $(0,0)$, standard ghosts, antighosts and auxiliary fields will have, respectively, bidegrees $(1\,,\,0)\,,\,(0\,,\,1)\,(1\,,\,1)\,$.

An interesting point to be remarked is that the bidegree distinguishes  objects that have the same ghost number but different roles in the extended algebra. A simple example is,
taking a classical gauge field, the associated BRST antifields $(-1\,,\,0)$ and the antighosts $ (0\,,\,1)\,$  or, similarly, the anti-BRST antifields $(0\,,\,-1)$ and the ghosts $(1\,,\,0)$.  

In order to build up this kind of duplicated BRST  formulation in a form that is completely symmetrical in the BRST and anti-BRST sectors, in the Lagrangian field antifield case\cite{GH3}, one needs to introduce two antibrackets, one with bidegree $(1\,,\
,0)$ and the other $(0\,,\,1)\,$.
Following these lines of (duplicated) standard BRST cohomology, but always keeping this symmetry between the two sectors represented in the bidegree, one arrives at the same results of \cite{BLT1}.

In order to gauge fix one needs  extra sets of auxiliary fields $\pi_A^{a}$ 
\cite{BLT1}  that  are introduced in \cite{GH3} as conjugated to the ${\bar \phi}^{A}$ in the corresponding antibracket (they are represented as  $\mu^A_a\,$ in  this reference).
One can then show that gauge fixing is equivalent to that of standard field antifield quantization, for a particular kind of gauge fixing fermions (see also \cite{H2,BLT3}).

 \section{Triplectic quantization}

A detailed description of triplectic quantization can be found
in refs. \cite{T1,T2,T3,DJB}. We will just make a short review. Considering some gauge theory, we enlarge the original field content, adding all the usual gauge fixing structure: ghosts, antighosts and auxiliary fields associated with the original gauge s
ymmetries. The resulting set will be denoted as $\,\phi^A\,$. Then we associate with each of these fields five new quantities, introducing the sets:  $ {\bar \phi}^A $, $\,\phi_A^{\ast\,1}\,$ ,
$\,\phi_A^{\ast\,2}\,$,$\,\pi_A^{1}\,$ and  $\pi_A^{2}\,$. The Grassmanian parities of these fields are: 
$\epsilon (\phi^A) \,=\,\epsilon ({\bar \phi}^A) \,\equiv \,\epsilon_A\,$,
$\epsilon (\phi^{\ast\,a}_A ) \,=\,\epsilon ( \pi^a_A ) \,=\,\epsilon_A\,+\,1\,$.
In this 6n dimensional space one defines
the two kinds of antibrackets ($\,a\,=\,1\,,\,2\,$)

\begin{equation}
\label{AB}
\{ F\,,\,G\,\}^a\,\equiv \,{\partial^r F\over \partial \phi^A} 
{\partial^l G \over \partial \phi_A^{\ast\,a}}\,+\,
{\partial^r F\over \partial {\bar \phi}^A} 
{\partial^l G \over \partial \pi_A^a\,}
\,-\, {\partial^r F \over \partial \phi_A^{\ast\,a}}\,
{\partial^l G\over \partial \phi^A}\,-\,
{\partial^r F \over \partial \pi_A^a\,}{\partial^l G\over \partial {\bar \phi}^A} 
\end{equation}

\noindent and also introduces a triplectic generalization of the $\Delta $ operator

\begin{equation}
\Delta^a\,\equiv\, (-1)^{\epsilon_A}\,
 \,{\partial^l \over \partial \phi^A} 
{\partial^l  \over \partial \phi_A^{\ast\,a}}\,+\,(-1)^{\epsilon_A}\,
{\partial^l \over \partial {\bar \phi}^A} 
{\partial^l  \over \partial \pi_A^a\,}
\end{equation}

\noindent and the operators

\begin{equation}
V^a\,=\,{1\over 2} \epsilon^{ab}\,\Big( \phi_A^{\ast\,b}
{\partial^r \over \partial {\bar \phi}^A}  
- (-1)^{\epsilon_A} \pi_A^b {\partial^r \over \partial \phi^A} \Big)\,\,.
\end{equation}

\noindent here and in the rest of the article, unless explicitly indicated, we are
adopting the convention of summing over repeated indices.

The quantum action $\,W\,$ is a solution of the two master equations:

\begin{equation}
\label{ME}
{1\over 2} \{ W \,,\, W\,\}^a \,+\,V^a W \,=\,
i\hbar \Delta^a W
\end{equation}

The Vacuum functional is defined as

\begin{equation}
\label{VAC}
Z\,=\, \int [{\cal D}\phi][\Dc \phi^{\ast}][\Dc \pi]
[\Dc \bar\phi ][\Dc \lambda]\,exp\{{i\over \hbar}\big(
W\,+\,X\big)\}
\end{equation}
 
\ni
where the functional $X$, that depends on extra fields $\lambda^A\,$, will represent gauge fixing and must satisfy the  equations

\begin{equation}
{1\over 2} \{ X \,,\, X\,\}^a \,-\,V^a X \,=\,
i\hbar \Delta^a X
\end{equation}

\ni where the sign of the $V^a$ term is the opposite as  that of the master equations (\ref{ME}).

For a gauge theory with closed and irreducible algebra,   
corresponding to a classical action $S_0[\phi^i]$, a solution for the action $W$ at classical level is:

\begin{equation}
\label{TAC}
S_{Tr.} \,=\, S_0 + \phi_A^{\ast\,a}\,{\overline \delta}_a \phi^A
+ {1\over 2} {\bar \phi}_A {\overline \delta}_2 {\overline \delta}_1 \phi^A 
\,+\,{1\over 2} \epsilon^{ab} \phi_A^{\ast\,a}\,\pi^A_b
\end{equation}

\ni where the ${\overline \delta}_a $ represent  gauge fixed BRST ($a=1$) and anti-BRST ($a=2$) transformations of the fields.

\section{Canonical transformations in the \break triplectic quantization}

In the standard (only BRST invariant) BV formalism, one has just one antibracket, corresponding to a simpler version of the equation (\ref{AB}) without the terms involving $\pi$ and $\bar \phi$ and without the index $a$. 
Canonical transformations, in this case, are transformations in the field antifield space such that calculating the antibracket of two quantities and then transforming the result gives the same outcome as transforming the quantities  and then calculating 
the antibracket.
Therefore, the antibrackets between the fundamental objects
(fields and antifields) are not changed by canonical transformations.
These transformations   can be expressed in terms of a fermionic  generator of ghost number $-1$ depending on the old fields and the new antifields $F\,[\phi^A\,,\,\phi^{\ast\,\prime}_A\,]$

\begin{eqnarray}
\label{CBV}
\phi^{\prime\,A} &=& {\partial F \over \partial \phi^{\ast\,\prime}_A}
\no\\
\phi^{\ast\,A}&=& {\partial F \over \partial \phi_A}
\end{eqnarray}

\noindent If the matrix

\begin{equation}
\label{15}
M^{AB}\,=\, {\partial^r \partial^r F \over \partial \phi^{\ast\,\prime}_A \partial \phi_B}\,
\end{equation}

\ni is invertible, transformation (\ref{CBV}) is canonical\cite{BV3,TPN}.

\bigskip

Let us consider now the triplectic quantization. For each of the antibrackets  of eq. (\ref{AB}) with $a=1,2$ we can introduce a generator 
$F_a\,[\phi^A\,,{\bar \phi}^A\,,\phi^{\ast\,a\,\prime}_A\,,
\pi^{a\,\prime}_A\,\,]\,$ and write out the set of transformations

\begin{eqnarray}
\label{CTR1}
\phi^{A\,\prime} &=& {\partial F_a \over \partial \phi^{\ast\,a\,\prime}_A}
\no\\
\phi_A^{\ast\,a} &=& {\partial F_a \over \partial \phi^A}\no\\
{\bar \phi}^{A\,\prime} &=& {\partial F_a \over \partial \pi^{a\,\prime}_A}
\no\\
\pi_A^{a} &=& {\partial F_a \over \partial {\bar \phi}^A}\,,
\end{eqnarray}

\ni where there is no sum over $a$. If the matrix

\begin{equation}
\label{invers}
 T_a^{\alpha\,\beta} \,=\,{ \partial^r \partial^r  F_a \over \partial z^{\ast\,a\,\prime}_\alpha \partial z_\beta }
\end{equation}

\ni (where there is again no sum over $a$ and we are defining $\{ z^\alpha \} \,\equiv \, \{ \phi^A \,,\,{\bar \phi}^A \} \,$ and 
$\, \{ z^{\ast \, a\,\prime}_\alpha \} \,\equiv\,\{ \phi^{\ast\,a\,\prime}_A \,,\,
\pi^{a\,\prime}_A \}\,\,$) is invertible, each of  these transformations, for fixed $a=1$ or $2$ will not change  the form of the corresponding antibracket.

 Now let us consider two generators $F_1$ , $F_2$  both with non singular matrices (\ref{invers}) but satisfying also the additional constraints

\begin{eqnarray}
\label{conditions}
{\partial F_1 \over \partial \phi^{\ast\,1\,\prime}_A} &=&
{\partial F_2 \over \partial \phi^{\ast\,2\,\prime}_A}
\no\\& &\no\\
{\partial F_1 \over \partial \pi^{1\,\prime}_A} &=&
{\partial F_2 \over \partial \pi^{2\,\prime}_A}\,\,.
\end{eqnarray}

\bigskip

\ni In this case the complete set of transformations  (\ref{CTR1}) including both $a=1$ and  $a=2$ will  leave
 the two antibrackets invariant, preserving the complete triplectic anticanonical structure.
We will therefore call  them as canonical transformations in the triplectic space.

The constraints (\ref{conditions}) restrict the possible dependence of the generators of these transformations on the variables $\phi^{\ast\,a\,\prime}_{A}$ and
$\pi^{a\,\prime}_A\,$. Their general form is 

\begin{equation}
F_a \,=\,{\bf 1}_a\,+\,f_a 
\end{equation}

\ni with

\begin{eqnarray}
{\bf 1}_a &=&  \phi^A \phi^{\ast\,\prime}_{A\,a}
\,+\,{\bar \phi}^A \pi^{\prime}_{A\,a} \no\\& &\no\\
f_1 &=& g_1 [\phi\,,{\bar \phi}]\,+\,g_3^A[\phi\,,{\bar \phi}] \pi^{ 1\,\prime}_A \,+\,
g_4^A [\phi\,,{\bar \phi}]\phi^{\ast\,1\,\prime}_A \no\\& &\no\\
f_2 &=&  g_2[\phi\,,{\bar \phi}] 
\,+\, g_3^A [\phi\,,{\bar \phi}] \pi^{2\,\prime}_A\,+\,
g_4^A [\phi\,,{\bar \phi}]  \phi^{\ast\,2\,\prime}_A \,\,,
\end{eqnarray}

\ni where we have  explicitly separated an identity operator ${\bf 1}_a\,$  just for future convenience.

\bigskip
Now going back to the equation (\ref{CTR1}) we see that general  triplectic canonical transformations can be put in the form

\begin{eqnarray}
\label{CTR2}
\phi^{\prime\,A} &=& \phi^{A}\,+\,g^A_4[\phi\,,{\bar \phi}]
\no\\
\phi^{\ast\,a\,}_A &=& \phi^{\ast\,a\,\prime}_A \,+\,
{\partial^r g_a \over \partial \phi^A}[\phi\,,{\bar \phi}]
\,+\,{\partial^r g_3^B \over \partial \phi^A} [\phi\,,{\bar \phi}] \pi_B^{a\,\prime}
\,+\,{\partial^r g_4^B \over \partial \phi^A}[\phi\,,{\bar \phi}]
\phi_B^{\ast\,a\,\prime}  \no\\
{\bar \phi}^{A\,\prime} &=& {\bar \phi}^A \,+\, 
g_3^A [\phi\,,{\bar \phi}]\no\\
\pi_A^{a} &=& \pi_A^{a\,\prime} \,+\,
 {\partial^r g_a \over \partial {\bar \phi}^A}[\phi\,,{\bar \phi}]
\,+\,{\partial^r g_3^B \over \partial {\bar \phi}^A}[\phi\,,{\bar \phi}] \pi_B^{a\,\prime}
\,+\,{\partial^r g_4^B \over \partial {\bar \phi}^A} [\phi\,,{\bar \phi}]
\phi_B^{\ast\,a\,\prime}  \no\\
\end{eqnarray}

\bigskip

An important point to be remarked is that the inversibility condition on the matrix 
$T_a^{\alpha\,\beta}$ of equation (\ref{invers}) does not imply the condition (as it happens in the usual BV) of inversibility of the matrix $M^{AB}$ of equation (\ref{15}).
 An interesting example to illustrate this point is the transformation generated by:

\begin{equation}
F^a\,=\,\phi^A \pi^{a\,\prime}_A \,+\,{\bar \phi}^A \phi^{\ast\,a\,\prime}_A
\end{equation}

\ni that clearly satisfies the  constraints (\ref{conditions}) and for which the matrix (\ref{invers}) is invertible but the matrix $M^{AB}\,$ is singular. The corresponding transformations are 

\begin{equation}
\pi^{a\,\prime}_A\,=\,\phi^{\ast\,a}_A\,\,,\,\,\,\,\phi^\prime_A\,=\,{\bar \phi}_A\,\,
\,\,,\,\,\,\,\phi^{\ast\,a\,\prime}_A\,\,=\,\,\,\,\pi^{\,a}_A\,\,\,,\,\,\,\,{\bar \phi}^{A\,\prime}\,=\,
\phi^A \,\,,
\end{equation}\bigskip

\ni that interchange $\phi \leftrightarrow {\bar \phi}$ , $ \phi^{\ast\,a}    \leftrightarrow \pi^a $. Therefore the antibrackets (\ref{AB}) are clearly  unchanged
and indeed the transformation is canonical.

\bigskip

It is worth remarking that the canonical transformations (\ref{CTR1}) do not preserve, in general, the bidegree defined on section (2). Actually the canonically transformed variables are in general non homogeneous elements of this bigrading.
That means, they may have non defined bidegree. This is  what happens in the case of the canonical transformations that will be used in order to fix the gauge in the next section.   However, the transformed variables, in this gauge fixing case, will be ho
mogeneous in the (standard) ghost number grading.

\section{Gauge fixing by canonical transformations}

The gauge fixing procedure consists in the construction of a non degenerated (gauge fixed) action $S_{GF}$ that belongs to the same cohomological class of the classical action and therefore describes the same physical observables\cite{HT}.
Gauge fixing in the case with extended BRST invariance has been studied in \cite{BLT1,BLT3,H2,GH1,GH3}. The most general gauge fixing with BRST and anti-BRST invariance was discussed, in the Hamiltonian framework, in \cite{GH1}. It was shown there that th
e gauge fixed action is not in general of the form $S_{GF}\,=\,S_0\,+\,\delta_2\delta_1 \chi$. We will however be concerned here just with the  Sp(2) symmetric case described in section (3) for which this result holds.

Let us consider the triplectic functional of equation (\ref{VAC}) with the introduction of an appropriate gauge fixing functional $X$. Integrating over $ {\bar \phi}^A $, $\,\phi_A^{\ast\,a}\,$ 
 and $\pi_A^{\,a}\,$, the ultimate result will be the exponential of $i/\hbar\,\,$ times a (non degenerated)  gauge fixed action  of the form

\begin{equation}
\label{TGFA}
S_{GF} \,=\, S_0[ \phi^i ] \,+\, {\overline \delta}_2 {\overline \delta}_1 {\bf \chi} [\phi^A]
\end{equation}  
 
\ni This action is (by construction) trivially invariant under  BRST and anti-BRST transformations.
The precise relation between the bosonic functional ${\bf \chi}$ and the  triplectic gauge fixing action $X$ of equation (\ref{VAC}) is not relevant for our purposes here. The question that we want to address is that of how to arrive at a gauge fixed acti
on of the form (\ref{TGFA}) by starting with the triplectic action (\ref{TAC}) and performing a canonical transformation.

In the standard BV quantization gauge fixing corresponds to replacing  the antifields according to

\begin{equation}
\label{SBV}
\phi^{\ast}_A\,=\,{\partial \Psi\,[\phi] \over \partial \phi^A}
\end{equation}

\ni where  $\Psi$ is a gauge fixing fermion.
It is well known that the same result (\ref{SBV})
can be obtained by performing a canonical transformation in  the (non gauge fixed) action like in eq. (\ref{CBV}) with $F \,=\, \phi^A \phi^{\ast\,\prime}_A \,+\,\Psi\,[\phi]\,$ 
and then removing the new antifields.
Therefore, one can interpret gauge fixing as the process of canonically transforming to a gauge fixed basis. In this case there is a trivial relation between the gauge fixing fermion and the corresponding canonical (fermionic) generator.

\bigskip

In the triplectic quantization, gauge fixing is not so trivial. Even for irreducible gauge theories with closed algebra, one can not interpret gauge fixing as simply the process of replacing the  antifields (or their extended triplectic version) according
 to relations like (\ref{SBV}). 
However, based on the idea of \cite{GH3} of introducing all the extended  variables in  conjugate pairs with respect to the antibrackets, we arrive at a natural question.
Can we get  action (\ref{TGFA}) starting with the triplectic action  (\ref{TAC}), performing a canonical transformation and then removing all the extra triplectic 
space variables  $ {\bar \phi}^A $, $\,\phi_A^{\ast\,a}\,$ 
 and $\pi_A^{\,a}\,$?  We will see that this is possible. The relation between the generators of the canonical transformations and the gauge fixing boson ${\bf \chi}$ will nevertheless  be not so trivial as in the standard BV case, reflecting the large fr
eedom in the gauge fixing procedure in the triplectic quantization. 

\bigskip

Starting with the non gauge fixed triplectic action $S_{Tr.}$ of eq. (\ref{TAC}), we will perform a canonical transformation as in (\ref{CTR2}). However, as we do not want to transform the fields $\phi^A$  we will choose 

\begin{equation}
\label{g4}
g^A_4\,=\,0.
\end{equation}

\ni Expressing the result in terms of the transformed fields and imposing the condition that we get  (\ref{TGFA}) when $ {\bar \phi}^{A\,\prime} $, $\,\phi_A^{\ast\,a\,\prime}\,$
and $\pi_A^{\,a\,\prime}\,$ are set to zero we get the general relation

\begin{equation}
\label{general}
{\partial f^{\prime}_a\over \partial \phi^A }{\overline \delta}_a \phi^A \,+\,
{1\over 2} g^{\prime}_3 {\overline \delta}_2 \,{\overline \delta}_1 \phi^A \,-\,
{1\over 2}\epsilon^{ab} {\partial f^{\prime}_a\over \partial \phi^A }
{\partial f^{\prime}_b\over \partial {\bar \phi}^A }\,\,=\,
 {\overline \delta}_2 {\overline \delta}_1 {\bf \chi} [\phi^A]\,\,,
\end{equation}

\ni where we are defining the primed functions as the corresponding function, written in terms of $\phi^A$ and 
$ {\bar \phi}^{A\,\prime} $, taken at $ {\bar \phi}^{A\,\prime}\,=\,0 $

\begin{equation}
f^{\prime}_a [\phi]\,=\, 
f_a [\,\phi\,,\,{\bar \phi}(\phi\,,
\,{\bar \phi}^{\prime}\,)\, 
]\vert_{_{{\bar \phi}^{\prime}\,=\,0}}
\end{equation} 

\ni and a similar definition for $\,g^{\prime}_i\,$.

Any solution of this equation, for a given ${\bf \chi}$ that appropriately fix the gauge, 
will represent a canonical transformation leading to a gauge fixed basis. Let us look at  some  solutions of (\ref{general}). We will consider three kinds of solutions that illustrate the large freedom in the choice of transformations that lead to the sam
e gauge fixed action. 

\bigskip

\ni (i)$\,\,\,$ Let us start with the particular case of  solutions for the generators $f$ that involve only the fields $\phi^A$.
This case corresponds to transforming only the antifields $\phi_A^{\ast\,a}$.  It is easy to see that  

\begin{eqnarray}
\label{case1}
g_1 &=& \alpha {\overline \delta}_2 \chi \no\\
g_2 &=& \beta {\overline \delta}_1 \chi 
\end{eqnarray}

\ni with $g_3^A$ is a solution of (\ref{general}) if the coefficients $\alpha$ and $\beta$
satisfy  $\,\,\beta\,-\,\alpha\,=\,1\,$. The particular case $\alpha\,=\,-1\,$ corresponds to  
a standard BV gauge fixing as in (\ref{SBV})
but with a particular gauge fixing fermion $\Psi\,=\,{\overline \delta}_2 \chi $ that leads to an Sp(2) invariant action.

Similarly, the case $\beta\,=\,1\,$ represents  anti-BRST version. In other words
it corresponds to the gauge fixing action that would show up if one builds up a BV formalism based on anti BRST symmetry and then chooses a gauge fixing fermion that  is a BRST variation: $\Psi\,=\,{\overline \delta}_1 \chi\,$.

\bigskip

\ni (ii)$\,\,\,$ Consider now generators $f$ involving $\phi^A\, $ and ${\,\bar \phi}^A$. This case has no parallel in  the standard FA quantization. The choice of generators (that includes the  solution (i) as a particular case)

\begin{eqnarray}
\label{case2}
g_1 &=& a_1 {\overline \delta}_2 \chi \,+\, 2 a_2 {\bar \phi}^A {\overline \delta}_2 \phi_A \no\\
g_2 &=& a_3 {\overline \delta}_1 \chi \,+\,2 a_4 {\bar \phi }^A 
 {\overline \delta}_1 \phi_A \no\\
g_3 &=& 0
\end{eqnarray}

\ni is a solution of equation (\ref{general}) if 

\begin{equation}
a_1 ( a_4\,-\,1) \,+\,a_3 (a_2\,+\,1) \,=\,1\,\,.
\end{equation}

In this case we are transforming the variables $\,\phi_A^{\ast\,a}\,$ and $\pi_A^{\,a}\,$.

\bigskip

\ni (iii)$\,\,\,$ Choosing now a canonical transformation changing all kinds of extended antifields, the transformation generated by 

\begin{eqnarray}
\label{case3}
g_1 &=& 2 b_1 {\overline \delta}_2 \chi\,+\,b_1 {\bar \phi}^A {\overline \delta}_2 \phi_A\no\\
g_2 &=& 2 b_2 {\overline \delta}_1 \chi\,+\,b_2  {\bar \phi}^A {\overline \delta}_1 \phi_A\no\\
g_3^A &=& b_3 {\partial \chi \over \partial \phi_A}
\end{eqnarray}

\ni with $b_3 \ne 0$ leads to the gauge fixed action (\ref{TGFA}) if 

\begin{equation}
\label{C3}
b_2\,-\,b_1\,+\,b_1 b_2 =\,{1\over 2} \,\,.
\end{equation}

\ni In this case all the triplectic variables, except $\phi^A\,$ are transformed.

\bigskip 

It is interesting to note that the gauge fixing considered in section (8) of reference \cite{GH3}, for the case of the variables $\phi_A^{\ast\,1}\,$ and ${\bar \phi}^{A}\,$  can be obtained from our canonical transformations by setting $g_1\,=\,g_2\,=0\,
$ and $g_3^A$ as in equation (\ref{case3}). Note that the anticanonical structure of this reference
is not exactly the same  as that of \cite{T1} used here. This explains why there is a difference in the case of the $\phi_A^{\ast\,2}\,$ variable.

\section{ Example: Yang Mills theory}

The classical action in this case is

\begin{equation}
\label{YMA}
S_0\,=\,\int d^4x \Big( - {1\over 4} Tr ( F^{\mu\nu} F_{\mu\nu} )\,\Big)
\end{equation}

In references \cite{Sp,BLT1} one finds a general procedure for building up the extended BRST algebra associated with  an irreducible gauge theory with closed algebra.
A convenient way of writing out the algebra of the Yang Mills theory, that corresponds just to 
a redefinition in the auxiliary fields of \cite{Sp,BLT1} is that of \cite{BD} 

\begin{eqnarray}
{\overline \delta}_1 A_\mu &=&  D_\mu c \no\\
{\overline \delta}_1 c &=& - {1\over 2}\  [c,c]_+\no\\
{\overline \delta}_1 \tilde c &=& - G \no\\
{\overline \delta}_1 G &=& 0  \no\\ 
\no\\ 
{\overline \delta}_2 A_\mu &=&   D_\mu 
\tilde c \no\\
{\overline \delta}_2 c &=&  \omega ( G - [c,\tilde c]_+ )
\no\\
{\overline \delta}_2\  \tilde c &=& -
{ 1 \over 2}\  [\tilde c,\tilde c]_+ \no\\
{\overline \delta}_2 G &=&  [ G, \tilde c ]  
\end{eqnarray}

\ni where   $\, c\,\equiv\,c^a T^a\,$,$\,{\tilde c} \,\equiv\,{\tilde c}^a T^a \,$ and $\,G\,\equiv\, G^a T^a\,$  are respectively the ghosts, antighosts and auxiliary fields.
We are using $Tr\,(\,T^a\,T^b\,)\,=\,{1 \over 2}$.

\bigskip

The triplectic action  for the Yang Mills case is just  expression (\ref{TAC})  where 
$\phi^A \,=\,\{ A_\mu , \, c\,,\,\tilde c \,,\,G\,\}$ and we associate with each of these fields the
corresponding extended antifields, that we will represent as 
$ {\bar \phi}^A =\,\{ {\bar A}_\mu \, {\bar c}\,,\,\bar {\tilde c} \,,\,\bar G \,\} $
, $\,\phi_A^{\ast\,a}\,=\,\{ A_\mu^{\ast\,a} \,,\, c^{\ast\,a}\,,\,{\tilde c}^{\ast\,a} \,,\,G^{\ast\,a}\,\}$ 
 and 
$ \pi_B^{a}\,=\,\{ \pi^{[A]\,a}_\mu \,,\,\pi^{[c]\,a}\,,\,\pi^{[ {\tilde c} ]\,a} \,,\,\pi^{[G]\,a}\,\}$. After integrating over all these extended antifields, we get again a result of the form  (\ref{TGFA}).

We will choose the same gauge fixing bosonic function as in ref.\cite{BD}

\begin{equation}
\label{YMB}
\chi\,=\,-{1\over 2} Tr \Big( A_\mu \,A^\mu\,\Big)
\end{equation}

\ni that corresponds to a gauge fixing action

\begin{equation}
\label{GFYM}
{\overline \delta}_2 {\overline \delta}_1 {\bf \chi} \,\,=\,\,
Tr \Big( \partial_\mu G  A^\mu \,+\,\partial_\mu {\tilde c} D^ \mu c \Big)
\end{equation}

\ni where $ D_\mu c\,=\, \partial_\mu c \,+ [ A_\mu\,,\,c\,]\,\,$.

Let us  look at  the canonical transformations that acting on

\begin{equation}
\phi_A^{\ast\,a}\,{\overline \delta}_a \phi^A
+ {1\over 2} {\bar \phi}_A {\overline \delta}_2 {\overline \delta}_1 \phi^A 
\,+\,{1\over 2} \epsilon_{ab} \phi_A^{\ast\,a}\,\pi_A^b
\end{equation}

\ni will generate (\ref{GFYM}) after removing the transformed (extended) antifields.
As discussed in the last section, the general condition is given by equation (\ref{general}) and there is a wide class of solutions. Let us  consider solutions corresponding to the three cases  discussed in the previous section.

\bigskip

\ni (i)$\,\,\,$  If we choose, for example, in the first case of equation (\ref{case1}) the parameter $\alpha\,=\,0$ the solution will be:

\begin{eqnarray}
\label{case1YM}
g_1 &=& 0  \no\\
g_2 &=& {\overline \delta}_1 \chi \,=\,Tr \Big( (\partial^\mu A_\mu ) c \Big)\no\\
g_3 &=& 0
\end{eqnarray}

\ni and the only fields that will transform are

\begin{eqnarray}
A^{\ast\,2\,\prime}_\mu &=& A^{\ast\,2 }_\mu\,+\,\partial_\mu c\no\\
c^{\ast\,2\,\prime} &=&  c^{\ast\,2}\,- \partial_\mu A^\mu
\end{eqnarray}

\bigskip

\ni (ii) $\,\,\,$ As an example of the second case, choosing in  equation (\ref{case2}) the parameters as $a_1\,= \,a_3\,=\,a_4\,=\, 1\,\,$, $\,\, a_2\,=\,0$,
we get the generators 

\begin{eqnarray}
f_1 &=& - Tr \,\Big[ A^\mu \partial_\mu {\tilde c} \Big]\no\\
f_2&=& - Tr \,\Big[ 2 {\bar A}^\mu D_\mu c 
\,-\,{\bar c} [\,c\,,\,c\,]_+ \,-\,2{\bar {\tilde c}} \chi \,-\,
 A^\mu \partial_\mu c \Big]\,\,.
\end{eqnarray}

\ni In this case the transformations in the fields will be

\begin{eqnarray}
A^{\ast\,1\,\prime}_\mu &=& A^{\ast\,1 }_\mu\,+\,\partial_\mu {\tilde c}\no\\
{\tilde c}^{\ast\,1\,\prime} &=&  {\tilde c}^{\ast\,1}\,- \partial_\mu A^\mu \no\\
A^{\ast\,2\,\prime}_\mu &=& A^{\ast\,2 }_\mu\,+\,\partial_\mu  c \no\\
 c^{\ast\,2\,\prime} &=&  c^{\ast\,2}\,+ 2 \partial_\mu {\bar A}^\mu
\,+2i \,[\, A_\mu\,,\,\bar A^\mu\, ]\,-\,\partial_\mu A^\mu  \no\\
 G^{\ast\,2\,\prime} &=& G^{\ast\,2}\,+ 2 {\bar {\tilde c}}
\end{eqnarray}

\bigskip

(iii) $\,\,\,$ The third case of last section, with arbitrary coefficients, and gauge fixing boson of the form of equation (\ref{YMB}) corresponds to the canonical transformation generated by:

\begin{eqnarray}
f_1 &=&  Tr \Big( - b_3 \,A^\mu  \pi^{[A]\,1\,\prime}_\mu \,+\, 2 \,b_1 (  {\bar A}^\mu
D_\mu {\tilde c }\,+\, {\bar c}\, G \,+\,  i {\bar c} \,[ \,{\tilde c}\,,\,c\,]_+ \,+\,
{i\over 2} {\bar {\tilde c}} \,[\, {\tilde c}\,,\,{\tilde c}\,]_+ \no\\
&-& i {\bar G}\,[ \,{\tilde c}\,,\, G\,]\,+\,\partial_\mu A^\mu \,{\tilde c}\,)\,
 \,\Big)\no\\
& &\no\\
f_2 &=&  Tr \Big(\,- b_3 A^\mu  \pi^{[A]\,2\,\prime}_\mu \,+\,2\,b_2 (  {\bar A}^\mu
\,D_\mu  c \,+\,{i \over 2}{\bar  c}\, [\, c\,,\, c\,]_+ \no\\
&-& \bar {\tilde c}\, G\,+ \,\partial_\mu A^\mu \, c \,)\,\, \Big)
\end{eqnarray}

\ni where, of course, condition (\ref{C3}) must be satisfied.

The corresponding transformations of the triplectic variables will be

\begin{eqnarray}
A_\mu^{\ast\,1\,\prime} &=& A_\mu^{\ast\,1}\,+\,{1\over 2} b_3 \pi^{[A]\,1}_\mu
\,+\, b_1\,\Big( -i [\,{\tilde c}\,,\,{\bar A}_\mu \,]\,+\,\partial_\mu {\tilde c} \Big) \no\\
& &\no\\
c^{\ast\,1\,\prime} &=& c^{\ast\,1}\,-\,i b_1\,[\,{\bar c}\,,\,{\tilde c}\,]_+\no\\
{\tilde c}^{\ast\,1\,\prime} &=& {\tilde c}^{\ast\,1}\,+\,b_1\, \Big( \partial^\mu {\bar A}_\mu \,+\,
i [\,\, A^\mu\,,\,{\bar A}_\mu \,]\,+\,i [\,c\,,\,{\tilde c}\,]_+
\no\\
&+& -\,{i\over 2}\,[\,\,{\tilde c}\,,\, \bar {\tilde c}\,]_+\,+\,i \,[ G\,,\,{\bar G}\,]
\,-\,\partial_\mu A^\mu \Big) \no\\
& & \no\\
G^{\ast\,1\,\prime} &=& G^{\ast\,1}\,-\,b_1\,{\bar c}
\end{eqnarray}

\begin{eqnarray}
A_\mu^{\ast\,2\,\prime} &=& A_\mu^{\ast\,2}\,+\,{1\over 2} b_3 \pi^{[A]\,2}_\mu
\,+\, b_2\, \Big( -i [\,c\,,\,{\bar A}_\mu \,]\,+\,\partial_\mu  c \Big) \no\\
& &\no\\
c^{\ast\,2\,\prime} &=& c^{\ast\,2}\,+\, b_2\,\Big( \partial^\mu {\bar A}_\mu \,
- i\, [\,{\bar A}_\mu \,,\, A^\mu\,]\,
-\,i\,[\,\bar c\,,\, c\,]_+\,+\,\partial_\mu A^\mu \Big)\no\\
{\tilde c}^{\ast\,2\,\prime} &=& {\tilde c}^{\ast\,2}\,  \no\\
G^{\ast\,2\,\prime} &=& G^{\ast\,2}\,+\,b_2 \bar {\tilde c}
\\
& & \no\\
{\bar A}_\mu^\prime &=& {\bar A}_\mu\,-\,{1\over 2} b_3\, A_\mu \,\no\\
{\bar c}^\prime &=& {\bar c} \, \no\\
{\bar {\tilde c}}^\prime &=& {\bar {\tilde c}}\no\\
{\bar G}^\prime &=& {\bar G} 
\end{eqnarray}

\begin{eqnarray}
\pi^{[A]\,1\,\prime}_\mu &=&  \pi^{[A]\,1}_\mu\,-\,b_1\,D_\mu {\tilde c}\no\\
\pi^{[c]\,1\,\prime} &=&    \pi^{[c]\,1}\,+\,b_1 \,\Big(\, G\, +\,i\, [ {\tilde c}\,,\,c\,]_+ \,\Big)\no\\
\pi^{[ {\tilde c}]\,1\,\prime} &=&   \pi^{[ {\tilde c}]\,1} \,+\,{i\over 2}\,b_1
\,[\,{\tilde c}\,,\,{\tilde c}\,]_+\no\\
\pi^{[G]\,1\,\prime} &=&   \pi^{[G]\,1} \,+\,i b_1 [\,{\tilde c}\,,\,G\,]
\end{eqnarray}

\begin{eqnarray}
\pi^{[A]\,2\,\prime}_\mu &=&  \pi^{[A]\,2}_\mu\,-\,b_2\,D_\mu c \no\\
\pi^{[c]\,2\,\prime} &=&    \pi^{[c]\,2}\,+\,{i \over 2}\,b_2[c\,,\,c\,]_+ \no\\
\pi^{[ {\tilde c}]\,2\,\prime} &=&   \pi^{[ {\tilde c}]\,2} \,-\,b_2 \,G \no\\
\pi^{[G]\,2\,\prime} &=&   \pi^{[G]\,2} 
\end{eqnarray}

\newpage

\section{Conclusions}
Canonical transformations have an important role in the field antifield  formalism.
In this article we found a general form for canonical transformations  in the extended field antifield space of the triplectic quantization.
We have shown that the condition of both  antibrackets been invariant places  constraints on the form of the generators, that have no analog in the standard BV space.
We have also  shown that it is possible to change the triplectic fields to a gauge fixed basis by means of these canonical transformations. The wide range of possibilities for transformations leading to the same gauge fixed action nicely reflects the larg
e freedom in the gauge fixing procedure for this Sp(2) invariant formalism.

\vskip 1cm
\section{Acknowledgements}
The authors are partially supported by CAPES, CNPq., FINEP and FUJB (Brazilian Research Agencies).
\vfill\eject

\end{document}